\documentclass{iopart}

\usepackage{epsfig}
\usepackage{bm}
\usepackage{epsfig}

\newcommand{\bd}{\bm}

\newcommand{\G}{\mathcal{G}}
\newcommand{\Z}{\mathcal{Z}}

\newcommand{\idf}{\int\!\!D\Phi\,\,}

\newcommand{\Go}{\mathbf{G}_0}
\newcommand{\p}[1]{\frac{\delta}{\delta #1}}
\newcommand{\pp}[2]{\frac{\delta #1}{\delta #2}}
\newcommand{\ppt}[3]{\frac{\delta^{(2)} #1}{\delta #2 \delta #3}}
\renewcommand{\Tr}{\mathrm{Tr}}

\begin{document}
\title[fRG with spontaneous symmetry breaking]{Functional renormalization group  
with vacuum expectation values and spontaneous symmetry breaking}

\author{Florian Sch\"{u}tz and Peter Kopietz}
\address{Institut f\"{u}r Theoretische Physik, Universit\"{a}t
  Frankfurt,  Max-von-Laue Strasse 1, 60438 Frankfurt, Germany}
\ead{schuetz@itp.uni-frankfurt.de}

\date{December 1, 2005}

\begin{abstract}
  We generalize our recently developed super-field functional
  renormalization group (RG) method involving both Fermi and Bose
  fields [F. Sch\"{u}tz, L. Bartosch, and P. Kopietz, Phys. Rev. B
  {\bf{72}}, 035105 (2005)] to include the possibility that some
  bosonic components of the field have a finite vacuum expectation
  value.  We derive an exact hierarchy of flow equations for the
  one-line irreducible vertices and the vacuum expectation value of
  the field. We apply our method to an interacting Fermi system where
  the interaction can be decoupled in the zero-sound channel and is
  then mediated by a collective bosonic field. The vacuum expectation
  value of the zero-frequency and zero-momentum component of the
  bosonic field is then closely related to the fermionic density.
  This can be exploited to calculate the compressibility of the
  interacting system. By using a cutoff in the bosonic propagator, the
  RG can be set up such that the self-consistent Hartree approximation
  is imposed as initial condition for the RG flow and the corrections
  to this approximation are generated as the remaining degrees of
  freedom are successively eliminated by the RG procedure.
\end{abstract}

\submitto{\JPA}


\section{Introduction}

Recently, we have proposed (in collaboration with L. Bartosch) a new
formulation of the functional renormalization group approach to
interacting Fermi systems which is based on the explicit introduction
of collective bosonic fields representing the relevant collective
fluctuations via a suitable Hubbard-Stratonovich transformation
\cite{Schuetz05}.  In contrast to the conventional renormalization
group (RG) approach to quantum critical phenomena pioneered by Hertz
\cite{Hertz76}, in our approach the Fermi fields are not completely
integrated out, but the RG flow of the coupled Fermi-Bose theory is
derived using the functional RG equation for the generating functional
of the one-line irreducible vertices \cite{Wetterich93,Morris94}.  A
similar strategy has recently been adopted by Baier, Bick and
Wetterich \cite{Baier04}. However, on the technical level our
formulation differs from that of Ref. \cite{Baier04}.  In particular,
we succeeded in introducing the RG cutoff parameter $\Lambda$ in such
a way that the asymptotic Ward identities underlying the exact
solubility of the one-dimensional Tomonaga-Luttinger model are
preserved.  This has enabled us to solve the infinite hierarchy of
flow equations of the coupled Fermi-Bose theory and reproduce the
exact solution for the single-particle Green function of the
Tomonaga-Luttinger model entirely within the framework of the RG.

As pointed out in Ref.~\cite{Schuetz05}, our method should be
especially convenient to study spontaneous symmetry breaking in
interacting Fermi systems provided that the order parameter can be
defined in terms of the vacuum expectation value of a suitable bosonic
field.  However, the exact RG flow equations in the form given in
Ref.~\cite{Schuetz05} are only valid as long as the fields do not have
finite vacuum expectation values.  We present here a generalization of
the RG flow equations derived in Ref.~\cite{Schuetz05} which
explicitly allows for finite vacuum expectation values of one or
several components of the fields.  Note that this does not necessarily
mean that some symmetry is spontaneously broken (see
Sec.~\ref{sec:fermions}).

Previously, several authors have studied spontaneous symmetry breaking
in interacting Fermi systems within the framework of the functional RG
\cite{Salmhofer04,Honerkamp05,Gersch05}.  However, in these works a
purely fermionic version of the functional RG has been used. In view
of the bosonic nature of the order parameter field in interacting
Fermi systems, a formulation where the flow of the vacuum expectation
value of the order parameter field appears explicitly in the exact
hierarchy of flow equations seems to be more advantageous, in
particular in systems where the fluctuations of the order parameter
field are controlled by a non-Gaussian fixed-point.

\section{Exact RG flow equations with finite vacuum expectation values}

\begin{figure}[p]    
  \centerline{\epsfig{file=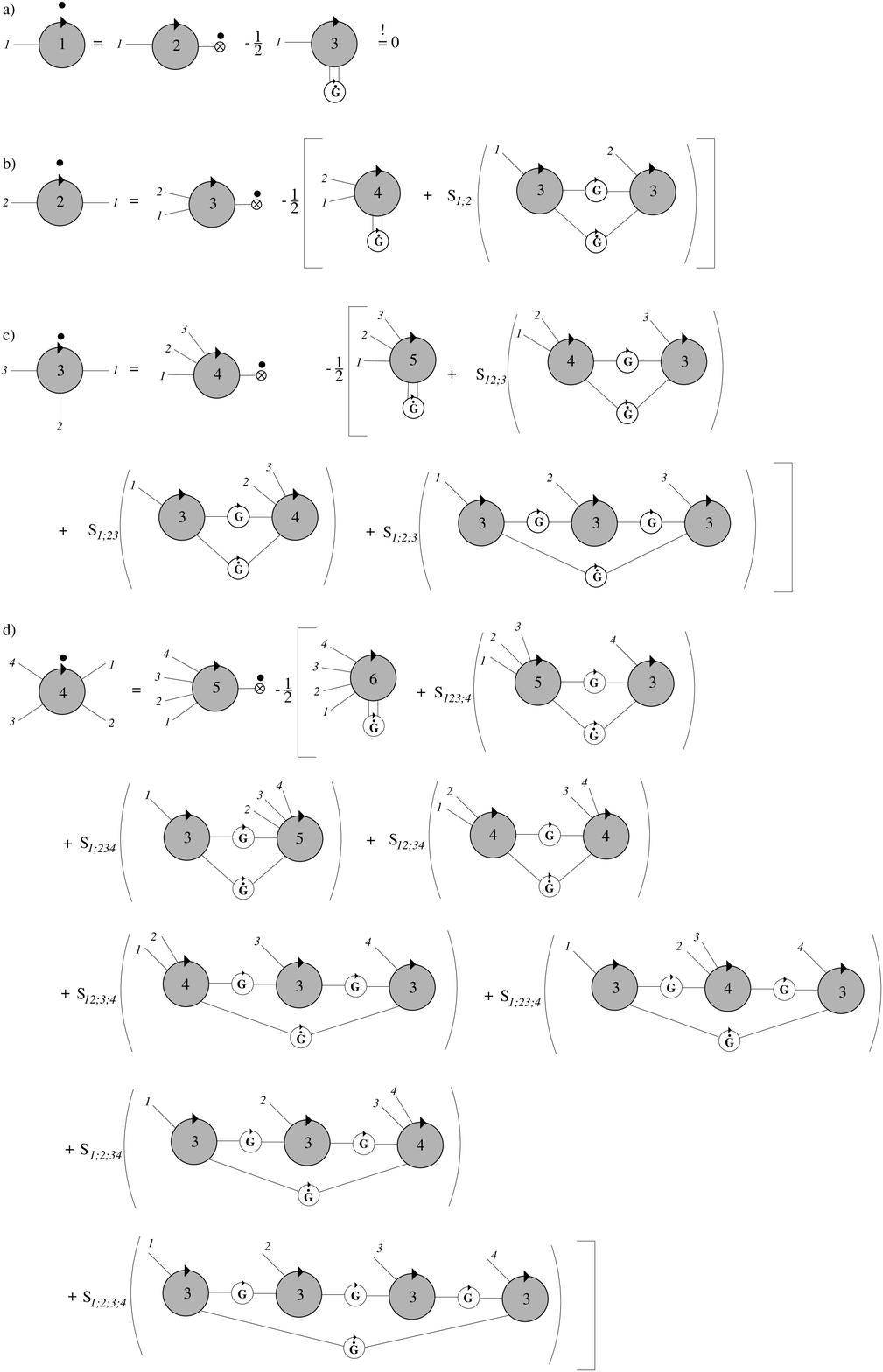,width=0.84\hsize}}
  \caption{%
    Diagrammatic representations of the exact RG flow equations for
    the first few one-line irreducible vertices: (a) $\Gamma^{(1)}$,
    (b) $\Gamma^{(2)}$, (c) $\Gamma^{(3)}$, (d) $\Gamma^{(4)}$. Shaded
    circles represent one-line irreducible vertex functions, whereas
    circles labeled by ${\bf G}$ and $\dot{\bf G}$ represent the full
    Green function and the single scale propagator defined in
    Eqs.~(\ref{eq:Dyson}) and (\ref{eq:Gdotmatrix}) respectively.  The
    small dotted circle with a cross denotes the change
    $\partial_{\Lambda} \Phi^0$ of the vacuum expectation value.  The
    action of the symmetrization operator $\mathcal{S}$ is defined in
    Eq.~(\ref{eq:symmopdef}).  The arrow on the circles indicates the
    order of the indices on the corresponding correlation function,
    which is important to keep track of the signs, when Fermi fields
    are present. }
    \label{fig:floweq}
  \end{figure}

In this section we consider a general Euclidean action $S [ \Phi ] =
S_0 [ \Phi] + S_1 [ \Phi ]$ involving some multi-component field
$\Phi_{\alpha}$ which can have fermionic and bosonic components.  The
index $\alpha$ is a super-label for all quantities that are necessary
to specify the fields, such as frequency, momentum, spin, and
field-type.  We write the quadratic part $S_0[ \Phi ]$ of the action
as
\begin{equation}
  S_0[\Phi]= 
  -\frac12\left(\Phi,\left[{\bf G}_0\right]^{-1}\Phi\right)
  = 
  - \frac12 \int_{\alpha} \int_{\alpha^{\prime}} \Phi_{\alpha}
  \left[{\bf G}_0\right]^{-1}_{\alpha \alpha^{\prime}} \Phi_{ \alpha^{\prime}}
  \,,
\end{equation}
where ${\bf G}_0$ is a matrix in all indices and the symbol
$\int_{\alpha}$ denotes integration over the continuous components and
summation over the discrete components of the super-index $\alpha$.
At this point it is not necessary to specify the interaction part $S_1
[ \Phi ]$, which can contain terms describing interactions between
more than two particles and couplings between the fermionic and
bosonic components of the super-field $\Phi_{\alpha}$.  The generating
functional $\G_c [J]$ of the connected Green functions can be defined
as the following functional integral,
\begin{equation}
  e^{\G_c[J]} = \frac{1}{\Z_0}\idf e^{-S_0-S_1+(J,\Phi)}
  \; ,
  \label{eq:Ggen}
\end{equation}
where $J_{\alpha}$ are super-sources associated with the super-fields,
and we use the compact notation
\begin{equation}
  (J,\Phi ) = \int_{\alpha} J_{\alpha} \Phi_{\alpha}
  \; .
  \label{eq:JPhiscalarproduct}
\end{equation}
The partition function $\Z_0$ of the non-interacting system can be
written as the Gaussian integral
\begin{equation}
  \Z_0 = \idf e^{-S_0}\,.
\end{equation}
Suppose now that at least one of the components of the super-field
$\Phi_{\alpha}$ has a finite expectation value (vacuum expectation
value) even for vanishing source fields $J$,
\begin{equation}
  \left. \frac{ \delta {\cal{G}}_c [ J ] }{ \delta J_{\alpha} }    \right|_{ J\to0}
  = \left. \langle \Phi_{\alpha}  \rangle \right|_{ J\to0}
  \equiv \Phi_{\alpha}^0 \neq 0
  \; .
\end{equation}
We now derive an exact hierarchy for RG flow equations for the
one-line irreducible vertices and for the vacuum expectation value.
The exact hierarchy obtained in \cite{Schuetz05} is only applicable if
none of the field components has a finite vacuum expectation value.
To generalize the hierarchy of flow equations, we proceed as
usual and introduce a cutoff $\Lambda$ into the theory by modifying
${\bf G}_0$ such that the long-wavelength and low-energy modes are
suppressed.  At this point it is not necessary to specify the precise
cutoff procedure.  Differentiating Eq.~(\ref{eq:Ggen}) with respect to
$\Lambda$ we obtain~\cite{Schuetz05}
\begin{equation}
  \fl \partial_{\Lambda} {\cal{G}}_c = 
  \frac12\left(
    \pp{\G_c}{J}, [ \partial_{\Lambda} \Go^{-1}]\pp{\G_c}{J}\right)
  +\frac12\Tr\left( [ \partial_{\Lambda} \Go^{-1}]\left[\ppt{\G_c}{J}{J}\right]^T\right)
  -\partial_{\Lambda}\ln\Z_0\,,
  \label{eq:Gcflow}
\end{equation}
where we have defined the following matrix in the super-indices,
\begin{equation}
  \left[\ppt{\G_c}{J}{J}\right]_{\alpha \alpha^{\prime}} =
  \frac{ \delta^{(2)} \G_c}{ \delta J_{\alpha} \delta J_{\alpha^{\prime}} }
  \; .
\end{equation}
To discuss symmetry breaking, it is more convenient to consider the
Legendre transform of ${\cal{G}}_c [ J ] $,
\begin{equation}
  {\cal{L}} [ \Phi ] = ( J [ \Phi ] , \Phi ) - {\cal{G}}_c [ J [ \Phi ]]
  \label{eq:Legendredef} 
\end{equation}
which has an extremum at $\Phi_{\alpha} = \Phi_{\alpha}^0$,
\begin{equation}
  \left. \frac{ \delta {\cal{L}} [ \Phi ] }{ \delta \Phi_{\alpha} }
  \right|_{ \Phi = \Phi^0} = 0
  \; . 
  \label{eq:Lextrem}
\end{equation}
Actually, the field $\Phi$ in Eqs.~(\ref{eq:Legendredef}) and
(\ref{eq:Lextrem}) denotes the expectation value of the field $\Phi$
in Eq.~(\ref{eq:Ggen}).  To simplify the notation, we use the same
symbol for both quantities; it is understood that from now on we
redefine $ \langle \Phi_{\alpha} \rangle \rightarrow \Phi_{\alpha}$.
In order to derive the exact RG flow equations of the one-line
irreducible vertices $\Gamma^{(n)}_{\alpha_1,\dots,\alpha_n} $ in the
presence of a finite vacuum expectation value $\Phi^0$, we consider
the flow equation for the generating functional
\begin{equation}
  \Gamma [ \Phi ] =  {\cal{L}} [ \Phi ] 
  + \frac{ 1}{2} ( \Delta \Phi , {\bf{G}}_0^{-1}
  \Delta \Phi )
  \; ,
  \label{eq:Gammadef}
\end{equation}
where $\Delta \Phi_{\alpha} = \Phi_{\alpha} - \Phi_{\alpha}^0$. Note
that by construction $\Gamma [ \Phi ] $ is also extremal at
$\Phi_{\alpha} = \Phi_{\alpha}^0$,
\begin{equation}
  \left. \frac{ \delta  \Gamma [ \Phi ] }{ \delta \Phi_{\alpha} }
  \right|_{ \Phi = \Phi^0} = 0
  \; .
  \label{eq:Gammaextrem}
\end{equation}
The one-line irreducible vertices
$\Gamma^{(n)}_{\alpha_1,\dots,\alpha_n} $, which depend implicitly on
the vacuum expectation value $\Phi^0$, are then defined via the
functional Taylor expansion around $\Phi^0$ (see, e.g.,
\cite{Berges02} and references therein),
\begin{equation}
  \Gamma[\Phi]=\sum_{n=0}^{\infty}\frac1{n!}\int_{\alpha_1}\dots
  \int_{\alpha_n}\Gamma^{(n)}_{\alpha_1,\dots,\alpha_n} \Delta \Phi_{\alpha_1}
  \cdot\ldots\cdot\Delta \Phi_{\alpha_n}
  \; .
  \label{eq:Gammaexpansion}
\end{equation}
Using Eqs.~(\ref{eq:Legendredef}), (\ref{eq:Gammadef}) and $\delta
{\cal{G}}_c / \delta J_{\alpha} = \Phi_{\alpha}$, we obtain the exact
flow equation
\begin{eqnarray}
  \fl\partial_{\Lambda} \Gamma = 
  -  \frac12\Tr\left( [ \partial_{\Lambda} \Go^{-1}]\left[\ppt{\G_c}{J}{J}\right]^T\right)
  + \partial_{\Lambda}\ln\Z_0 
  \nonumber\\
  -   ( \Delta \Phi , \partial_{\Lambda}[\Go^{-1}\Phi^0])
  - \frac{1}{2} ( \Phi^0 , [ \partial_{\Lambda} \Go^{-1} ] \Phi^0 )
  \; .
  \label{eq:Gammaflowprelim}
\end{eqnarray}
Introducing the matrices $\mathbf{\Sigma}$, $\mathbf{U}$, $\mathbf{G}$
and $\dot{\mathbf{G}}$ via
\begin{equation}
  \mathbf{\Sigma} =\left[\ppt{\Gamma}{\Phi}{\Phi} \right]_{\Phi=\Phi^0}^T
  \; ,
  \label{eq:Sigmadef}
\end{equation}
\begin{equation}
  \mathbf{U} =\left[\ppt{\Gamma}{\Phi}{\Phi}\right]^T
  -\left[\ppt{\Gamma}{\Phi}{\Phi}\right]_{\Phi=\Phi^0}^T
  =\left[\ppt{\Gamma}{\Phi}{\Phi}\right]^T-\mathbf{\Sigma}
  \; ,
  \label{eq:Udef}
\end{equation}
\begin{equation}
  \mathbf{G}^{-1}=\mathbf{G}_0^{-1}-\mathbf{\Sigma}\,,
  \label{eq:Dyson}
\end{equation}  
\begin{equation}
  \dot{\mathbf{G}} = - \mathbf{G}\partial_{\Lambda}[\mathbf{G}_0^{-1}]\mathbf{G}
  =[\mathbf{1}-\mathbf{G}_0\mathbf{\Sigma}]^{-1}
  \left(\partial_{\Lambda}\mathbf{G}_0
  \right)
  [\mathbf{1}-\mathbf{\Sigma}\mathbf{G}_0]^{-1}
  \; ,
  \label{eq:Gdotmatrix}
\end{equation}%
we may rewrite Eq.~(\ref{eq:Gammaflowprelim}) as follows,
\begin{eqnarray}
  \fl\partial_{\Lambda}\Gamma&=&
  -\frac12\Tr\left[\mathbf{Z}\dot{\mathbf{G}}^T\mathbf{U}^T
    \left\{
      \mathbf{1}-\mathbf{G}^T\mathbf{U}^T
    \right\}^{-1}\right]
  -   ( \Delta \Phi , \partial_\Lambda[\Go^{-1}\Phi^0])
  \nonumber\\[0.2cm]
  \fl&&-\frac12\Tr\left[\mathbf{Z}\dot{\mathbf{G}}_0^T\mathbf{\Sigma}^T
    \left\{
      \mathbf{1}-\mathbf{G}_0^T\mathbf{\Sigma}^T
    \right\}^{-1}\right]
  - \frac{1}{2} ( \Phi^0 , [ \partial_{\Lambda}\Go^{-1} ]  \Phi^0 )
  \; ,
  \label{eq:Gammaflow}
\end{eqnarray}
where $[{\bf Z}]_{\alpha \beta}=\delta_{\alpha,\beta}\zeta_{\alpha}$
with $\zeta_{\alpha}=\pm1$ depending on whether $\Phi_{\alpha}$ is a
bosonic or fermionic field.  To obtain the flow equations for the
vertices, we should take into account that $\Phi^0$ depends on
$\Lambda$, so that the derivative of Eq.~(\ref{eq:Gammaexpansion})
with respect to $\Lambda$ yields
\begin{eqnarray}
  \fl\partial_{\Lambda} \Gamma = 
  \sum_{n=0}^{\infty}\frac1{n!}\int_{\alpha_1}\dots
  \int_{\alpha_n}  \partial_{\Lambda} \Gamma^{(n)}_{\alpha_1,\dots,\alpha_n} \Delta \Phi_{\alpha_1}
  \cdot\ldots\cdot\Delta \Phi_{\alpha_n}
  \nonumber\\
  - \sum_{n=0}^{\infty}\frac1{n!} \int_{\alpha} \int_{\alpha_1}\dots
  \int_{\alpha_n}   \Gamma^{(n+1)}_{\alpha,\alpha_1,\dots,\alpha_n} 
  ( \partial_{\Lambda} \Phi^0_{\alpha} )
  \Delta \Phi_{\alpha_1}
  \cdot\ldots\cdot\Delta \Phi_{\alpha_n}
  \; .
  \label{eq:Gammaexpansionderiv}
\end{eqnarray}
Equating the right-hand sides of Eq.~(\ref{eq:Gammaflow}) and
(\ref{eq:Gammaexpansionderiv}) we obtain the flow equations for the
one-line irreducible vertices.  The vertex $\Gamma^{(0)}$ with no
external legs is the interaction correction to the grand canonical
potential. Its flow equation is
\begin{eqnarray}
  \fl\partial_{\Lambda}\Gamma^{(0)}=
  -  \frac12\Tr\left[\mathbf{Z}\dot{\mathbf{G}}_0^T\mathbf{\Sigma}^T
    \left\{
      \mathbf{1}-\mathbf{G}_0^T\mathbf{\Sigma}^T
    \right\}^{-1}\right]
  - \frac{1}{2} ( \Phi^0 , [ \partial_{\Lambda} \Go^{-1} ] \Phi^0 )
  - \!\!\int_{\alpha} ( \partial_{\Lambda} \Phi^0_{\alpha} ) \Gamma^{(1)}_{\alpha}
  .
  \label{eq:potflow}
\end{eqnarray}
The flow equation for the vertex $\Gamma^{(1)}_{\alpha} $ with one
external leg is
\begin{eqnarray}
  \fl \partial_{\Lambda} \Gamma^{(1)}_{\alpha} =
  - \frac{1}{2} 
  \Tr\left[\mathbf{Z}\dot{\mathbf{G}}^T \mathbf{\Gamma}^{(3)T}_{\alpha} \right]
  - \int_{\alpha'}  [ \mathbf{G}^{-1} ]_{ \alpha \alpha'} (\partial_{\Lambda} \Phi^0_{\alpha'})
  - \int_{\alpha'}  [ \partial_{\Lambda} \mathbf{G}_0^{-1} ]_{ \alpha \alpha'} 
  \Phi^0_{\alpha'}
  \; ,
  \label{eq:Gamma1flow}
\end{eqnarray}
and the flow equations for the vertices $\Gamma^{(n)}$ with $n \geq 2$
can be written in closed form as follows,
\begin{eqnarray} 
  \fl \partial_{\Lambda} {\Gamma}^{(n)}_{\alpha_1,\dots,\alpha_n} = 
  \int_{\alpha} ( \partial_{\Lambda} \Phi_{\alpha}^0  ) \Gamma^{(n+1)}_{ \alpha
    \alpha_1 \ldots \alpha_n }
  -\frac12\sum\limits_{l=1}^{\infty}
  \sum\limits_{m_1,\dots,m_l=1}^{\infty}\delta_{n,m_1+\ldots+m_l}\,
  \nonumber\\
  \times
  {\cal{S}}_{\alpha_1,\dots,\alpha_{m_1};\alpha_{m_1+1},\dots,\alpha_{m_1+m_2};
    \dots;\alpha_{m_1+\ldots+m_{l-1}+1},\dots,\alpha_{n}}
  \Big\{
  \nonumber\\
  \fl ~~~~~~~~\Tr\left[
    \mathbf{Z}\dot{\mathbf{G}}^T \mathbf{\Gamma}^{(m_1+2)\,T}_{\alpha_1,\dots,\alpha_{m_1}}\mathbf{G}^T
    \mathbf{\Gamma}^{(m_2+2)\,T}_{\alpha_{m_1+1},\dots,\alpha_{m_1+m_2}}
    \cdot\dots\cdot\mathbf{G}^T
    \mathbf{\Gamma}^{(m_l+2)\,T}_{\alpha_{m_1+\ldots+m_{l-1}+1},\dots,\alpha_{n}}
  \right] \Big\}
  \,,
  \label{eq:flow_vert}  
\end{eqnarray}
where the matrices $\mathbf{\Gamma}^{(n+2)}_{\alpha_1,\dots,
  \alpha_n}$ are defined as follows
\begin{equation}
  [ \mathbf{\Gamma}^{(n+2)}_{\alpha_1,\dots,
    \alpha_n} ]_{\alpha \alpha^{\prime} }
  = \Gamma^{(n+2)}_{ \alpha \alpha^{\prime}  \alpha_1 \dots \alpha_n }
  \; .
  \label{eq:Gammamatrix}
\end{equation}
The operator ${\cal{S}}$ symmetrizes the expression in curly brackets
with respect to indices on different correlation functions, i.e., it
generates all permutations of the indices with appropriate signs,
counting expressions only once that are generated by permutations of
indices on the same vertex. More precisely the action of ${\cal{S}}$
is given by ($m=\sum_{i=1}^lm_i$)
\begin{eqnarray}
  \fl {\cal{S}}_{\alpha_1,\dots,\alpha_{m_1};\dots;\alpha_{m-m_l+1},\dots,\alpha_{m}}
  \{ A_{\alpha_1,\dots,\alpha_m} \}
  =
  \frac{1}{\prod_i m_i!}\sum_P 
  \mathrm{sgn}_{\zeta}(P) \,
  A_{\alpha_{P(1)},\dots,\alpha_{P(m)}}\,,
  \label{eq:symmopdef}
\end{eqnarray}
where $P$ denotes a permutation of $\{1,\dots,m\}$ and
$\mathrm{sgn}_{\zeta}$ is the sign created by permuting field
variables according to the permutation $P$, i.e.,
\begin{equation}
  \Phi_{\alpha_1}\cdot\ldots\cdot\Phi_{\alpha_m} = 
  \mathrm{sgn}_{\zeta}(P) \,
  \Phi_{\alpha_{P(1)}}\cdot\ldots\cdot\Phi_{\alpha_{P(m)}}
  \; . 
\end{equation}
We now adjust the flowing vacuum expectation value $\Phi^{0}$ such
that for any value of the cutoff $\Lambda$ the vertex
$\Gamma^{(1)}_{\alpha}$ vanishes. In this case $\Phi^0$ can be
identified with the flowing extremum of the effective potential.  The
above equations (\ref{eq:potflow}) and (\ref{eq:Gamma1flow}) can be
further simplified if we choose the inverse free propagator
${\bf{G}}_0^{-1}$ such that
\begin{equation}
  {\bf{G}}_0^{-1}  \Phi^0 =0
  \; , \qquad\; 
  [ \partial_{\Lambda} {\bf{G}}_0^{-1} ] \Phi^0 =0
  \; .
  \label{eq:Gcond}
\end{equation}
We assume that this can always be achieved, if necessary by including
appropriate counter-terms in the Gaussian part of the action.  Then
the last two terms on the right-hand sides of Eqs.(\ref{eq:potflow})
and (\ref{eq:Gamma1flow}) vanish and the latter equation becomes
\begin{equation}
  \int_{\alpha}  [ \mathbf{\Sigma}]_{ \alpha_1 \alpha} \partial_{\Lambda} \Phi^0_{\alpha}
  = \frac{1}{2} 
  \Tr\left[   
    \mathbf{\Gamma}^{(3)}_{\alpha_1}     \dot{\mathbf{G}}   \right]
  \; ,
  \label{eq:phiflow}
\end{equation}
which relates the RG flow of the vacuum expectation value
$\Phi^0_{\alpha}$ to the irreducible self-energy $\mathbf{\Sigma}$ and
the three-legged vertex $\Gamma^{(3)}$.  Diagrammatic representations
of the flow equations for $\Gamma^{(n)}$ with $n=1,2,3,4$ are shown in
Fig.~\ref{fig:floweq}.

\section{Flow equations in the symmetry broken phase of the Ising model}
\label{sec:Ising}

Before applying the general equations derived above to quantum systems
involving both fermions and bosons, it is instructive to examine these
equations in the context of classical $\varphi^4$-theory in $D$
dimensions, which describes the critical properties of the
$D$-dimensional Ising model. In this case our super-field $\Phi$ has
only a single component $\varphi$ and the super-index $\alpha$ labels
momenta $\bd{k}$ (or, alternatively, points $\bd{r}$ in
$D$-dimensional space).  Our initial action is $S [ \varphi ] = S_0 [
\varphi ] + S_1 [ \varphi ]$, where
\begin{equation}
  S_0 [\varphi ] = \frac{1}{2} \int_{\bd{k}} G_{0}^{-1} ( \bd{k} ) 
  \varphi_{ - \bd{k}} \varphi_{ \bd{k}}
  \; ,
  \label{eq:S0def}
\end{equation}
and the interaction part is
\begin{eqnarray}
  \fl S_1 [\varphi ] = \frac{r_0}{2} \int_{\bd{k}}  
  \varphi_{ - \bd{k}} \varphi_{ \bd{k}}
  + \frac{u_0}{ 4 !} \int_{ \bd{k}_1} \ldots \int_{ \bd{k}_4}
  ( 2 \pi )^D \delta ( \sum_{i=1}^4 \bd{k}_i  )    \varphi_{\bd{k}_1 } \cdots \varphi_{\bd{k}_4 } 
  \; .
  \label{eq:S1def}
\end{eqnarray}
Here $\int_{\bd{k}} =\frac1{V}\sum_{\bm k}\to \int \frac{ d^D k}{ (2
  \pi )^D}$, for $V\to\infty$, where $V$ is the volume of the system. Furthermore, using
a sharp cutoff in momentum space for simplicity, the cutoff-dependent
free propagator is given by
\begin{equation}
  G_0 ( \bd{k} )  = \frac{ \Theta ( \Lambda < | \bd{k} | < \Lambda_0 )}{ c_0 \bd{k}^2 }
  \; ,
\end{equation}
where $\Theta ( \Lambda < | \bd{k} | < \Lambda_0 )$ is unity if
$\Lambda < | \bd{k} | < \Lambda_0$ and vanishes otherwise.  For the
Ising model on a $D$-dimensional hypercubic lattice with lattice
spacing $a$ the bare parameters are $c_0 = ( 2 D )^{-1}$ and $r_0 =
a^{-2} ( T - T_c^{\rm MF} ) / T_c^{\rm MF}$, where $T_c^{\rm MF} $ is
the mean-field result for the critical temperature.  In the symmetry
broken phase the $\bd{k}=0$ Fourier component of the field has a
finite expectation value. It is then natural to shift
\begin{equation}
  \varphi_{\bd{k} } = \varphi^0_{\bd{k}}  + \Delta \varphi_{\bd{k} }
  \; , \qquad
  \varphi^0_{\bd{k}} =
  ( 2 \pi )^D \delta ( \bd{k} ) M
  \label{eq:varphishift}
  \; .
\end{equation}
For the Ising model, the parameter $M$ has the physical meaning of the
magnetization per unit volume.  Note that $G_0^{-1} ( \bd{k} )
\varphi^0_{\bd{k}} =0$, so that the condition (\ref{eq:Gcond}) is
satisfied.  This is the reason why we have included the quadratic part
of the action proportional to $r_0$ into the definition of the
interaction term in Eq.~(\ref{eq:S1def}).  Our initial action then
takes the form $S [\varphi ] = S_0 [ \Delta \varphi ] + S_1
[\varphi^0; \Delta \varphi]$ with
\begin{eqnarray}
  \fl S_1 [ \varphi^0; \Delta \varphi ] & = & S_1 [ \varphi \rightarrow \varphi^0 + 
  \Delta \varphi  ] =
  V \Bigr[ \frac{r_0}{2} M^2 + \frac{u_0}{ 4 !} M^4 \Bigr]
  \\\fl
  & + &  \Gamma^{(1)}_0 \Delta \varphi_0   +  \frac{1}{2} \int_{\bd{k}} \Gamma^{(2)}_0  ( \bd{k} )
  \Delta \varphi_{ - \bd{k}} \Delta \varphi_{ \bd{k}}
  \nonumber
  \\\fl
  & + & \frac{1}{3!} \int_{\bd{k}_1}    \int_{\bd{k}_2}  \int_{\bd{k}_3}   
  ( 2 \pi )^D     \delta (  \sum_{ i =1}^3 \bd{k}_i ) 
  \Gamma^{(3)}_0  ( \bd{k}_1  , \bd{k}_2 , \bd{k}_3 )
  \Delta \varphi_{  \bd{k}_1 } \Delta \varphi_{ \bd{k}_2 } \Delta \varphi_{ \bd{k}_3 } 
  \nonumber
  \\\fl
  &+ & \frac{1}{4!} \int_{\bd{k}_1}    \ldots  \int_{\bd{k}_4}   
  ( 2 \pi )^D \delta (  \sum_{ i =1}^4 \bd{k}_i )
  \Gamma^{(4)}_0  ( \bd{k}_1  , \bd{k}_2 , \bd{k}_3 , \bd{k}_4 )
  \Delta \varphi_{  \bd{k}_1 } \Delta \varphi_{ \bd{k}_2 } 
  \Delta \varphi_{ \bd{k}_3 } \Delta \varphi_{ \bd{k}_4 }
  \; , \nonumber
\end{eqnarray}
where the bare vertices now depend on $M$,
\begin{eqnarray}
  \Gamma^{(1)}_0 & = & r_0 M + \frac{u_0}{ 6 } M^3
  \label{eq:Gamma1initial}
  \; ,
  \\
  \Gamma^{(2)}_0  ( \bd{k} ) & = & r_0 + \frac{u_0}{2} M^2
  \label{eq:Gamma2initial}
  \; ,
  \\
  \Gamma^{(3)}_0  ( \bd{k}_1  , \bd{k}_2 , \bd{k}_3 ) & = & u_0 M
  \label{eq:Gamma3initial}
  \; ,
  \\
  \Gamma^{(4)}_0  ( \bd{k}_1  , \bd{k}_2 , \bd{k}_3 , \bd{k}_4 ) & = & u_0 
  \label{eq:Gamma4initial}
  \; .
\end{eqnarray}
We now require that $\Gamma^{(1)}_0 =0$, so that the initial value $M_0$
of the order parameter in our functional RG is the minimum of the
effective potential in the Landau approximation,
\begin{equation}
  M_0 = \left\{ \begin{array}{cc}
      0 & \mbox{for $ r_0 > 0 $}  \\
      \sqrt{ - 6 r_0 / u_0 } & \mbox{for $ r_0 < 0 $}
    \end{array}
  \right.
  \; .
  \label{eq:M0landau}
\end{equation}
Note that for $r_0 < 0$ we have
\begin{equation}
  \Gamma^{(2)}_0  ( \bd{k} )  =  \frac{u_0}{3} M_0^2
  \label{eq:Gamma2initial2}
  \; ,
\end{equation}
so that in this case the interaction part of our initial action can be
written in real space as
\begin{equation}
  S_1 [ \varphi ] = \int d^{D}r \frac{u_0}{ 4 ! } [ \varphi^2 ( \bd{r} ) - M_0^2 ]^2
  \; .
  \label{eq:S1init}
\end{equation}
With a sharp cutoff in momentum space we obtain from
Eq.~(\ref{eq:potflow}) for the interaction correction to the grand
canonical potential,
\begin{equation}
  \partial_{\Lambda} \Gamma_{\Lambda}^{(0)} = - \frac{V}{2}
  \int_{\bd{k}} \delta ( | \bd{k} | - \Lambda )
  \ln \left[ \frac{ c_0 \Lambda^2 + \Gamma^{(2)}_{\Lambda} ( \bd{k} ) }{ c_0 \Lambda^2 }
  \right]
  \; .
  \label{eq:flowgamma0}
\end{equation}
The order parameter flow equation (\ref{eq:phiflow}) reduces to
\begin{equation}
  ( \partial_{\Lambda} M_{\Lambda} ) \Gamma_{\Lambda}^{(2)} (0) 
  = -\frac{1}{2} \int_{\bd{k}} \dot{G}_{\Lambda} ( \bd{k} )
  \Gamma^{(3)}_{\Lambda} ( \bd{k} , - \bd{k} , 0 )
  \; .
  \label{eq:orderflow}
\end{equation}
For a sharp cutoff in momentum space, we have
\begin{equation}
  {G}_{\Lambda} ( \bd{k} ) = \frac{ \Theta ( \Lambda < | \bd{k} | < \Lambda_0 )}{ 
    c_0 \bd{k}^2 + \Gamma^{(2)} ( \bd{k} ) }
  \; ,
\end{equation}
and hence
\begin{equation}
  \dot{G}_{\Lambda} ( \bd{k} ) =- \frac{ \delta ( | \bd{k} | - \Lambda )}{ c_0 \Lambda^2 +
    \Gamma^{(2)} ( \bd{k} ) }
  \; .
\end{equation}
From Eq.~(\ref{eq:flow_vert}), we obtain for the two-point vertex
\begin{eqnarray}
  \fl\partial_{\Lambda} \Gamma^{(2)}_{\Lambda} ( \bd{k} )
  =  \frac{1}{2} \int_{ \bd{k}^{\prime} }
  \dot{G}_{\Lambda} ( \bd{k}^{\prime} ) \Gamma^{(4)}_{\Lambda} (
  \bd{k}^{\prime} , - \bd{k}^{\prime} , \bd{k} , - \bd{k} )
  \nonumber
  \\
  - \int_{ \bd{k}^{\prime} }
  \dot{G}_{\Lambda} ( \bd{k}^{\prime} ) 
  {G}_{\Lambda} ( \bd{k}^{\prime} + \bd{k} ) 
  \Gamma^{(3)}_{\Lambda} (
  \bd{k} , - \bd{k} - \bd{k}^{\prime} , \bd{k}^{\prime} )
  \Gamma^{(3)}_{\Lambda} (
  - \bd{k}^{\prime} , \bd{k} + \bd{k}^{\prime} , - \bd{k} )
  \nonumber
  \\
  +
  ( \partial_{\Lambda} M_{\Lambda} ) \Gamma_{\Lambda}^{(3)} ( \bd{k} , - \bd{k} , 0 )
  \; .
  \label{eq:flowGamma3}
\end{eqnarray} 
The above system of flow equations is exact but not closed.  In order
to obtain a closed system for the two-point vertex and the order
parameter, it is necessary to express the vertices $\Gamma^{(3)}$ and
$\Gamma^{(4)}$ in terms of $\Gamma^{(2)}$ and $M$.  We now propose a
simple polynomial truncation which is motivated by the generalized gradient
expansion for the effective potential \cite{Berges02}.  Guided by our
initial conditions (\ref{eq:Gamma3initial}), (\ref{eq:Gamma4initial})
and (\ref{eq:Gamma2initial2}), we truncate the above equations by
substituting on the right-hand sides
\begin{eqnarray}
  \Gamma^{(2)}_{\Lambda} ( \bd{k} ) & \approx & \frac{u_\Lambda}{3} M_{\Lambda}^2
  \label{eq:G2trunc}
  \; ,
  \\
  \Gamma^{(3)}_{\Lambda} ( \bd{k}_1  , \bd{k}_2 , \bd{k}_3 ) & \approx & u_\Lambda M_{\Lambda}
  \label{eq:G3initialtrunc}
  \; ,
  \\
  \Gamma^{(4)}_\Lambda  ( \bd{k}_1  , \bd{k}_2 , \bd{k}_3 , \bd{k}_4 ) & \approx & u_\Lambda 
  \label{eq:G4initialtrunc}
  \; .
\end{eqnarray}
If we ignore the momentum-dependencies of the vertices also on the
left-hand sides of the flow equations, this truncation amounts to
approximating
\begin{equation}
  \Gamma_{\Lambda} [ \varphi ] = 
  \int d^{D}r \frac{u_\Lambda}{ 4 ! } [ \varphi^2 ( \bd{r} ) - M_\Lambda^2 ]^2
  \; ,
  \label{eq:Gammainit}
\end{equation}
which is nothing but the quartic approximation to the effective
potential in the symmetry broken phase to zeroth order in the gradient
approximation \cite{Berges02}.  In this approximation we obtain from
Eq.~(\ref{eq:orderflow}) for the flow of the order parameter,
\begin{equation}
  \partial_{\Lambda} M_\Lambda^2 = - 3 \int_{\bd{k}} \dot{G}_{\Lambda} ( \bd{k} )
  \label{eq:flowM}
  \; ,
\end{equation}
while Eq.~(\ref{eq:flowGamma3}) reduces to
\begin{equation}
  \partial_{\Lambda} u_{\Lambda} = - 3 u_{\Lambda}^2 \int_{\bd{k}}
  \dot{G}_{\Lambda} ( \bd{k} ) G_{\Lambda} ( \bd{k} )
  \; .
  \label{eq:flowu}
\end{equation}
We shall not further analyze these equations, because they are
completely equivalent to the RG flow equations discussed by Berges et
al. \cite{Berges02} using the truncated effective potential
(\ref{eq:Gammainit}) to zeroth order in the gradient expansion.
Whether or not the polynomial truncation is justified is a different
issue; according to Refs. \cite{Parola93,Bonanno04}, this truncation
fails to describe some important features of the symmetry breaking
phenomenon.
With the help of Eq.~(\ref{eq:flowGamma3}) one can easily go beyond
this approximation. For example, we may solve this equation
iteratively by substituting the truncation in
Eqs.~(\ref{eq:G2trunc})--(\ref{eq:G4initialtrunc}) only on the
right-hand sides, retaining the momentum-dependence generated by the
Green's functions \cite{Busche02,Ledowski04}. From the
momentum-dependent self-energy we can then obtain an estimate for the
anomalous dimension.  Having gained some confidence in our RG
equations, we now focus on a more interesting problem involving both
bosonic and fermionic degrees of freedom.

\section{RG approach to interacting Fermions via partial bosonization
in the zero-sound channel}
\label{sec:fermions}

In the approach developed in \cite{Schuetz05}, tad-pole diagrams of
the Hartree-type have not been considered due to the assumed overall
charge neutrality of the system. For more general interactions, these
diagrams do contribute and can lead to a non-vanishing expectation
value of the field describing density fluctuations. In this section,
this effect is treated with the formalism developed above for symmetry
breaking.

We consider a system with a density-density interaction. After a
Hubbard-Stratonovich transformation in the zero-sound channel the
interaction is mediated by a real bosonic field $\varphi$ and the
resulting action reads \cite{Schuetz05}
\begin{equation}
  S[\bar{\psi},\psi,\varphi] = S_0 [\bar{\psi}, \psi] + S_0 [\varphi]+
  S_{1}[\bar{\psi}, \psi,\varphi]
  \; ,
  \label{eq:Spsiphidef}
\end{equation}
with the free parts
\begin{eqnarray}
  S_0 [\bar{\psi}, \psi]
  = -\sum_{\sigma}\int_{K} \bar{\psi}_{K\sigma}
  G^{-1}_{0,\sigma}(K)
  \psi_{K\sigma}
  \label{eq:S0psidef}\\
  S_0 [\varphi] =
  \frac12\sum_{\sigma\sigma'}\int_{\bar{K}} 
  [F_0(\bar{K})^{-1}]_{\sigma\sigma'}
  \varphi^*_{\bar{K}\sigma} \varphi_{\bar{K}\sigma'} 
  \label{eq:S0phi}
  \; ,
\end{eqnarray}
and the coupling between Fermi and Bose fields
\begin{equation}
  \fl S_{1}[\bar{\psi} , \psi,\varphi] =
  i\sum_{\sigma}\int_{{K}} \int_{\bar{K}}
  \bar{\psi}_{ K + \bar{K}, \sigma} \psi_{K \sigma} \varphi_{\bar{K}\sigma}
  + \frac1{2\beta V}\sum_{\sigma,\sigma'} [f_0^{-1}]^{\sigma\sigma'}\varphi_{0\sigma}\varphi_{0\sigma'}
  \,.
  \label{eq:Spsiphi}
\end{equation}
We use composite frequency-momentum indices $K=(i\omega,{\bd k})$ and
$\bar{K}=(i\bar{\omega},{\bar{\bd k}})$ which include fermionic and
bosonic momenta and Matsubara frequencies. The integration measure is
$\int_K = \frac{1}{\beta V}\sum_{\omega, {\bd{k}}}$ and reduces to
$\int\frac{d\omega}{2\pi}\frac{d^Dk}{(2\pi)^D}$ in the
zero-temperature and thermodynamic limit $\beta,V\to\infty$. The
propagators are given by
\begin{equation}
  G_{0,\sigma}(K)= \frac{\Theta(\Lambda_F < 
    |\xi_{{\bd k}\sigma}-\Sigma_{\sigma}^{*}(i0,{\bd k}_F)|/v_0 < \Lambda_{F,0})}
  {i\omega - \xi_{{\bd k}\sigma}}\,,
\end{equation}
\begin{equation}
  F_{0,\sigma\sigma'}(\bar{K}) = [f^{-1}_{\bar{\bd k}}(1-\delta_{\bar{K},0}/(\beta V))]^{-1}_{\sigma\sigma'}
  \Theta(\Lambda_B < |\bar{k}| < \Lambda_{B,0})\,,	
\end{equation}
with an arbitrary fermionic dispersion $\xi_{{\bd
    k}\sigma}=\epsilon_{{\bd k}\sigma} -\mu$ measured relative to the
chemical potential $\mu$, the Fourier transformed interaction
$f_{\bar{\bd k}}^{\sigma\sigma'}$, and some average Fermi velocity
$v_0$ which is introduced to give $\Lambda_F$ units of momentum. Here,
$\Sigma^{*}=\Sigma^{\Lambda_F=\Lambda_B=0}$ is the self energy of the
fully interacting system, which has to be introduced as a counter-term
in order to scale toward the true Fermi surface \cite{Kopietz01}.  We
have allowed for the possibility of independent cutoffs on bosonic and
fermionic degrees of freedom.

The Legendre transformation is performed for fermionic as well as
bosonic fields, such that the resulting functional
$\Gamma[\bar{\psi},\psi,\varphi]$ generates one-line irreducible
vertices that are irreducible in the particle propagator as well as
 interaction lines \cite{Schuetz05}. At finite density, the bosonic
field $\varphi$ acquires an expectation value and should be shifted
according to $\varphi_{\bar{K}\sigma} = \varphi^0_{\bar{K}\sigma} +
\Delta\varphi_{\bar{K}\sigma}$ with
\begin{equation}
  \varphi^0_{\bar{K}\sigma}= -i \delta_{\bar{K},0} \bar{\varphi}_{\sigma}\,,
  \qquad
  \bar{\varphi}_{\sigma}= \sum_{\sigma'} f_{0}^{\sigma\sigma'} \rho_{\sigma'}\,.
  \label{eq:rhodef}
\end{equation}
By integrating out the bosonic fields, it is straightforward to see
that for $\Lambda_F=\Lambda_B=0$, the quantity $\rho_{\sigma}$ is the
density of the interacting electron system in the spin channel
$\sigma$, i.e., $\rho_{\sigma}=\int_{\bd k} \langle\hat{c}_{{\bd
    k}\sigma}^{+}\hat{c}_{{\bd k}\sigma}\rangle$, where $\hat{c}_{{\bd
    k}\sigma}$ is the usual second quantized fermion annihilation
operator.  The zero frequency and momentum parts of the interaction
have not been included in $F_0$ to fulfill the property
(\ref{eq:Gcond}) of the free propagators.  The functional Taylor
expansion
\begin{eqnarray}
  \fl \Gamma[\bar{\psi}, \psi,\varphi] = \sum_{n=0}^{\infty}\sum_{m=0}^{\infty}
  \frac{1}{(n!)^2m!}
  \int_{K_1'\sigma_1'}\dots\int_{K_n'\sigma_n'}
  \int_{K_1\sigma_1}\dots\int_{K_n\sigma_n}
  \int_{\bar{K}_1\bar{\sigma}_1}\dots\int_{\bar{K}_m\bar{\sigma}_m}
  \nonumber \\[1mm]
  \times
  \delta_{ K'_1 + \ldots + K'_n ,  K_1 + \ldots + K_n  + \bar{K}_1 + \ldots + \bar{K}_m }
  \nonumber \\[1mm]
  \times
  \Gamma^{(2n,m)}(K'_1\sigma_1',\dots,K'_n\sigma_n';
  K_1\sigma_1,\dots,
  K_n\sigma_n;\bar{K}_1\bar{\sigma}_1,\dots,\bar{K}_m\bar{\sigma}_m) 
  \nonumber\\[1mm]
  \times
  \bar{\psi}_{K'_1\sigma_1'}\cdot\ldots\cdot\bar{\psi}_{K'_n\sigma_n'}
  \psi_{K_1\sigma_1}\cdot\ldots\cdot\psi_{K_n\sigma_n}
  \Delta\varphi_{\bar{K}_1\bar{\sigma}_1}\cdot\ldots\cdot\Delta\varphi_{\bar{K}_m\bar{\sigma}_m}\,,
  \label{eq:expansion2}
\end{eqnarray}
defines the vertex functions $\Gamma^{(2n,m)}$.
	
\subsection{Fermi surface cutoff scheme}

In the standard RG approach to interacting Fermi systems
\cite{Shankar94}, the fermionic degrees of freedom are integrated out
in successive shells around the (interacting) Fermi surface.  In our
formulation this is achieved by setting $\Lambda_B=0$ and by taking
only $\Lambda=\Lambda_F$ as a running cutoff. In this case our general
flow Eq.~(\ref{eq:phiflow}) reduces to the following flow equation for
the expectation values $\bar{\varphi}_{\sigma}$:
\begin{equation}
  \sum_{\sigma'}\Pi_{\sigma\sigma'}(0)(\partial_{\Lambda}\bar{\varphi}_{\sigma'}) = 
  i\zeta\int_{K} \dot{G}_{\sigma}(K)\Gamma^{(2,1)}(K\sigma;-K\sigma;0\sigma)\,,
  \label{eq:phibarflow}
\end{equation}
where $\zeta=-1$. The fermionic self-energy satisfies the flow
equation
\begin{eqnarray}
  \partial_{\Lambda}\Sigma_{\sigma}(K) &=& -i\,\Gamma^{(2,1)}(K\sigma;K\sigma;0\sigma)
  (\partial_{\Lambda}\bar{\varphi}_{\sigma})\nonumber\\
  &+& \int_{K'} \dot{G}_{\sigma}(K')F_{\sigma\sigma'}(K-K')
  \Gamma^{(2,1)}(K'\sigma;K\sigma;K'-K,\sigma)\nonumber\\
  &&~~~~~~~~~~~~~~~~~\times\Gamma^{(2,1)}(K\sigma;K'\sigma;K-K',\sigma)
  \nonumber\\
  &-& \zeta \sum_{\sigma'}\int_{K'} \dot{G}_{\sigma'}(K')
  \Gamma^{(4,0)}(K\sigma,K'\sigma';K'\sigma',K\sigma)\,,
  \label{eq:sigmaflow}
\end{eqnarray}
where for a sharp cutoff, the propagators are given by
\begin{eqnarray}
  \dot{G}_{\sigma} (K) =  - \frac{ \delta ( \Lambda - |\xi_{{\bd k}\sigma}+
  \Sigma^{*}_{\sigma}(i0,{\bd k}_F)|/v_0 ) }
  {i \omega - \xi_{ {\bf{k}} \sigma} - \Sigma_{\sigma}(K)}
  \; ,
  \label{eq:dotGdiagdef}
  \\
  G_{\sigma}(K) = \frac{\Theta(\Lambda < 
 |\xi_{{\bd k}\sigma}+\Sigma_{\sigma}^{*}(i0,{\bd k}_F)|/v_0 < \Lambda_{0})}
	{i\omega - \xi_{{\bd k}\sigma} - \Sigma_{\sigma}(K)}\,,
  \label{eq:Gdiagdef}
  \\
  F_{\sigma \sigma^{\prime}} ( \bar{K} ) =  
  \left[F_{0}(\bar{K})^{-1}  + \Pi(\bar{K})\right]^{-1}_{\sigma,\sigma^{\prime}}
  \; .
  \label{eq:Fdiagdef}
\end{eqnarray}
On the right-hand side of the last equality it is understood, that the
$\Theta$-function in $F_{0}$ should be omitted.  The fermionic
self-energy and the polarization are defined by
$\Sigma_{\sigma}(K)=\Gamma^{(2,0)}(K\sigma;K\sigma)$, and
$\Pi_{\sigma\sigma'}({\bar{K}})=\Gamma^{(0,2)}(-\bar{K}\sigma,\bar{K}\sigma')$
respectively.

To explore these flow equations in a simple situation, we will now
neglect the flow of $\Gamma^{(2,1)}$, $\Gamma^{(4,0)}$ as well as
$\Pi$ and set these vertices equal to their initial values. This
approximation then reduces to an RG version of the Hartree-Fock
approximation.  No frequency dependencies of the self-energy are
generated, i.e., $\Sigma_{\sigma}(K)=\Sigma_{\sigma}({\bd k})$, and we
can thus define a flowing quasi-particle dispersion
$\xi^{\Lambda}_{{\bd k}\sigma}=\xi_{{\bd
    k}\sigma}+\Sigma_{\sigma}({\bd k})$. Using Eqs.~(\ref{eq:rhodef})
and (\ref{eq:phibarflow}), we find that the flow of the density is
given by
\begin{equation}
  \fl \partial_{\Lambda}\rho_{\sigma} =
  -\int_{\bd{k}} \delta(|\xi_{{\bd k}\sigma}
  +\Sigma_{\sigma}^{*}(i0,{\bd k}_F)|/v_0-\Lambda)
  n_F (\xi^{\Lambda}_{{\bd k}\sigma})\,,
  \label{eq:rhoflowsimple}
\end{equation}
while Eq.~(\ref{eq:sigmaflow}) determines the flow of the quasi-particle
dispersion,
\begin{equation}
  \fl \partial_{\Lambda}\xi_{{\bd k},\sigma}^{\Lambda} = 
  \sum_{\sigma'}f_0^{\sigma\sigma'}
  (\partial_{\Lambda}\rho_{\sigma'})
  +\int_{ \bd{k}^{\prime}} \delta(|\xi_{{\bd k'}\sigma}
  +\Sigma^{*}_{\sigma}(i0,{\bd k}_F)|/v_0-\Lambda)
  f^{\sigma\sigma}_{{\bd k}-{\bd k}'} n_F (\xi^{\Lambda}_{{\bd k}'\sigma})\,,
  \label{eq:xiflow}
\end{equation}
where $n_F( \epsilon )=[e^{\beta \epsilon }+1]^{-1}$ is the Fermi
function, and we have explicitly set $\zeta=-1$.  To further simplify
these equations, we use a Hubbard-type interaction with $f_{\bar{\bd
    k}}^{\sigma\sigma'}= f_{0} \delta_{\sigma,-\sigma'}$. In this case
no extra momentum dependencies of the quasi-particle dispersion are
generated, so that we may set $\Sigma_{\sigma} ( \bd{k} ) = - \Delta
\mu_{\sigma}$.  Our counter-term is then simply
$\Sigma^{*}_{\sigma}(i0,{\bd k}_F) = - \lim_{\Lambda \rightarrow 0}
\Delta \mu_{\sigma}$, and the flowing quasi-particle dispersion
can be written as $\xi_{{\bf k}\sigma}^{\Lambda}=\xi_{{\bm
    k}\sigma}-\Delta\mu_{\sigma}$.  If the bare dispersion $\xi_{{\bm
    k}\sigma}$ is actually independent of the spin projection
$\sigma$, then also $\Delta \mu_{\sigma} = \Delta \mu$ and
$\rho_{\sigma} = \rho$ are spin-independent.  In this case $\Delta
\mu$ can be interpreted as the flowing interaction correction to the
chemical potential.  The flow equation (\ref{eq:xiflow}) for the
energy dispersion then reduces to
\begin{eqnarray}
  \partial_{\Lambda}\Delta\mu = 
  - f_0(\partial_{\Lambda}\rho)
  \; .
\end{eqnarray}
Integrating this equation, we find that, for a given value of the
cutoff $\Lambda$, the relation between the flowing interaction
correction $\Delta \mu$ to the chemical potential and the flowing
density is simply
 \begin{eqnarray}
  \Delta\mu  = - f_0\rho\,.
\end{eqnarray}
Substituting this result into
Eq.~(\ref{eq:rhoflowsimple}), we obtain a closed flow equation for the RG flow
of the density $\rho$ of electrons (per spin projection),
\begin{eqnarray}
  \partial_{\Lambda}\rho =
  - v_0[\nu_0(\mu-f_0\rho^*+v_0\Lambda) n_F ( f_0  (\rho-\rho^*)+v_0\Lambda)
  \nonumber\\~~~~~~~~~~~
  +\nu_0(\mu-f_0\rho^*-v_0\Lambda) n_F (f_0(\rho-\rho^*)-v_0\Lambda)]\,,
 \label{eq:rhoflow}
\end{eqnarray}
where  $\rho^{\ast} = \rho_{ \Lambda =0}$ is the true density 
(per spin projection) and
\begin{equation}
  \nu_{0}(\epsilon) \equiv \int_{\bd{k}}
  \delta(\epsilon_{{\bd k}\sigma}-\epsilon)
\end{equation}
is the density of states of the non-interacting system.
At zero temperature, Eq.~(\ref{eq:rhoflow}) can
easily be integrated to yield an implicit relation between the true
$\rho^*$ and the chemical potential $\mu$, 
\begin{equation}
  \rho^*= \int_{A}^{\mu-f_0\rho^*}d\epsilon\,\nu_0(\epsilon)\,,
\end{equation}
where $A=\mu-f_0\rho^*-v_0\Lambda_0$ is the lower band edge. From this 
expression, we obtain the usual Fermi-liquid result for the compressibility 
\cite{Pines89}, 
\begin{equation}
  \kappa = \frac{\partial \rho^*}{\partial\mu} = \frac{\nu_0}{1+f_0\nu_0}\,,
\end{equation}
where $\nu_0=\nu_0(\mu - f_0 \rho^*)$ is the density of states at the
true Fermi surface.

\subsection{Interaction cutoff flow}

Instead of the standard cutoff procedure for interacting fermions
\cite{Shankar94} discussed in the previous section, we have proposed
in Ref. \cite{Schuetz05} to use a running cutoff in the bosonic sector
of the theory. Such a cutoff procedure was also used in the original
work by Hertz~\cite{Hertz76} on quantum critical phenomena in
interacting electron systems.  However, in contrast to our approach,
Hertz completely eliminated the fermionic degrees of freedom.  In the
notation used here, we implement the interaction-cutoff procedure by
setting $\Lambda_F=0$ and by using $\Lambda=\Lambda_B$ as a running
cutoff.
\begin{figure}[t]
  \begin{center}
    \epsfig{file=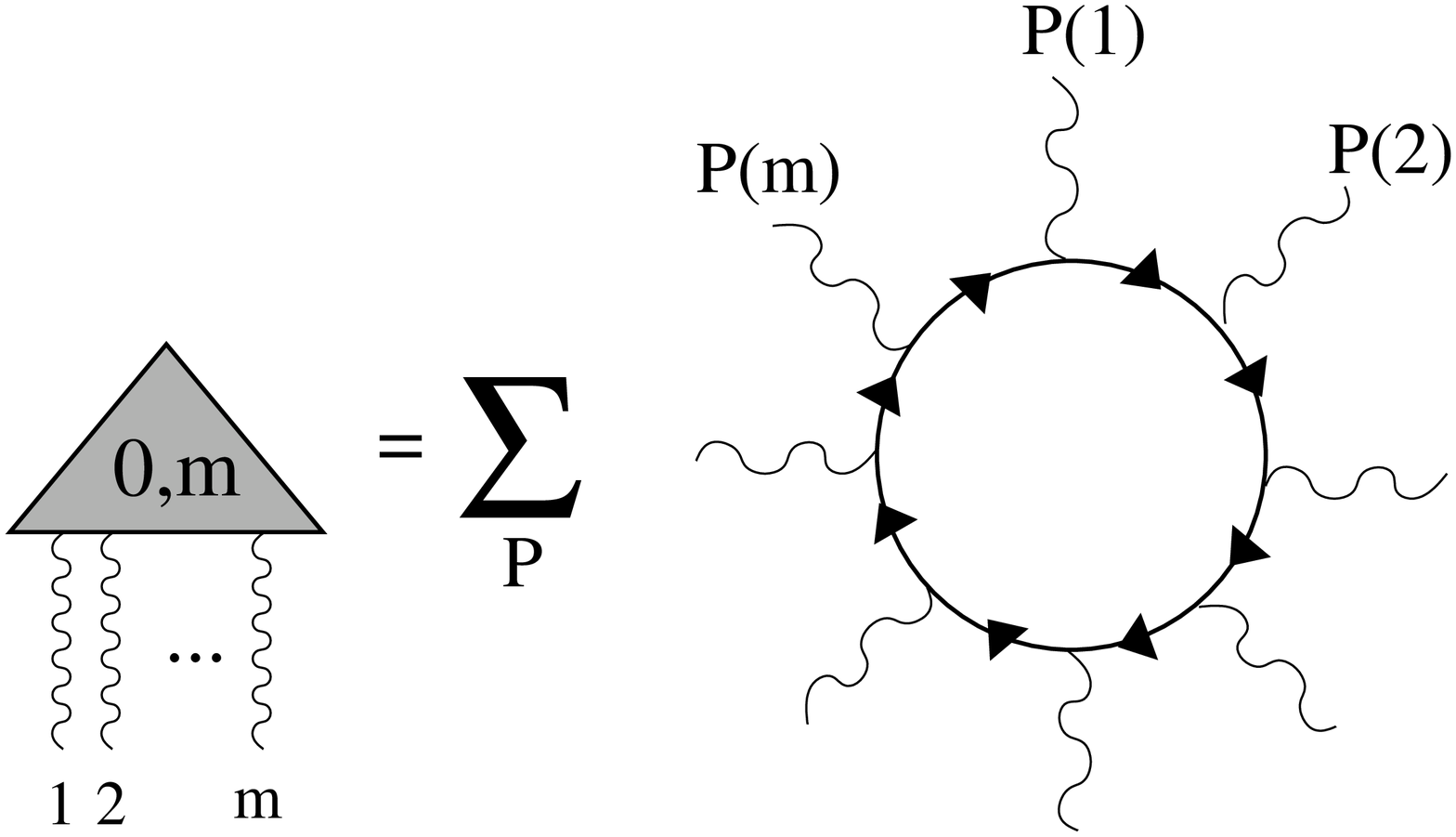,width=0.7\hsize}
  \end{center}
  \caption{ Initial condition for the pure boson vertices in the
    momentum transfer cutoff scheme. The sum is taken over the $m!$
    permutations of the labels of the external legs.}
  \label{fig:closedloop}
\end{figure}
The initial condition for the flow in this interaction cutoff scheme
corresponds to a situation in which interaction lines are turned off
while particle propagator lines are fully functional. In addition to
the bare three-leg interaction vertex, the only one-line irreducible
diagrams that can be drawn are closed loops of fermionic propagators.
These loops have to be symmetrized with respect to the exchange of
external bosonic legs in order to obtain the initial condition for the
purely bosonic vertices as shown in Fig.~\ref{fig:closedloop}. A more
formal derivation of the initial condition is given in the Appendix.
The result is given by
\begin{equation}
  \fl \Gamma_{\Lambda_0}[\bar{\psi},\psi,\varphi] =
  i (\bar{\psi},\hat{\varphi}\psi) 
  +\frac1{2\beta V}\sum_{\sigma\sigma'} [f_0^{-1}]^{\sigma\sigma'} 
  \varphi_{0\sigma}\varphi_{0\sigma'}
  - \Tr \{ \ln[\hat{1}-i\hat{G}_0\hat{\varphi}] \}\,,
  \label{eq:Gamma0}
\end{equation}
where the matrix $\hat{\varphi}$ of bosonic fields is defined in
analogy to Eq.~(\ref{eq:Jmatrixdef})
and 
\begin{equation}
  \fl [\hat{{G}}_0]_{K\sigma,K'\sigma'} = \delta_{K,K'}\delta_{\sigma,\sigma'}
{G}_{0,\sigma}(K)\,,
  \qquad
  {G}_{0,\sigma}(K) = [i\omega - {\xi}_{{\bd k}\sigma}]^{-1}\,.
\end{equation}
The first two terms in Eq.~(\ref{eq:Gamma0}) are nothing but the bare
interaction $S_1$ defined in Eq.~(\ref{eq:Spsiphi}), whereas the last
term generates the closed fermion loops.

When expanded in powers of the field $\varphi$, Eq.~(\ref{eq:Gamma0})
contains linear terms. We thus need to perform the shift
$\varphi_{\bar{K}\sigma} = \varphi_{\bar{K}\sigma}^0+
\Delta\varphi_{\bar{K}\sigma}$ with $\varphi^0_{K \sigma}$ given in
Eq.~(\ref{eq:rhodef}), yielding
\begin{eqnarray}
  \fl \Gamma_{\Lambda_0}[\bar{\psi},\psi,\varphi] = 
  -\Tr \{ \ln[\hat{1}-i\hat{G}_{\varphi}\Delta\hat{\varphi}] \}
  +\frac1{2\beta V}\sum_{\sigma\sigma'} [f_0^{-1}]^{\sigma\sigma'} 
  \Delta\varphi_{0\sigma}\Delta\varphi_{0\sigma'}
  -i \sum_{\sigma} \Delta\varphi_{0\sigma} \rho_{\sigma}
  \nonumber\\
  \fl~~~~~~~~~~~~~~+\bar{\varphi}_{\sigma}(\bar{\psi},\psi)
  + i (\bar{\psi},\Delta\hat{\varphi}\psi)
  - \Tr \{ \ln[\hat{G}_0\hat{G}_{\varphi}^{-1}]\} 
  -\frac{\beta V}{2}\sum_{\sigma\sigma'} f_0^{\sigma\sigma'}\rho_{\sigma}\rho_{\sigma'} \,,
  \label{eq:Gammafto0}
\end{eqnarray}
where we have defined the shifted bare Green function
\begin{equation}
  \hat{G}_{\varphi}^{-1} = \hat{G}_0^{-1} -i\hat{\varphi}^0\,,
\end{equation}
with the matrix elements
\begin{equation}
  \fl [\hat{G}_{\varphi}]_{K\sigma,K'\sigma'} = 
  \delta_{K,K'}\delta_{\sigma,\sigma'}\tilde{G}_{0,\sigma}(K)\,,
  \qquad
  \tilde{G}_{0,\sigma}(K) = [i\omega - \tilde{\xi}_{{\bd k}\sigma}]^{-1}\,,
\end{equation}
and $\tilde{\xi}_{{\bd k}\sigma} \equiv \xi_{{\bd k}\sigma} +
\sum_{\sigma'}f_0^{\sigma\sigma'}\rho_{\sigma'}$.  From
Eq.~(\ref{eq:Gammafto0}) the initial condition for the
one-line irreducible vertices can be obtained by expanding the
right-hand side in powers of $\Delta\varphi$ and by symmetrizing the
resulting expressions with respect to the interchange of bosonic
fields. We obtain
\begin{eqnarray}
  \Gamma^{(0,0)}_{\Lambda_0} = -\Tr \{\ln[\hat{G}_0\hat{G}_{\varphi}^{-1}]\}
  -\frac{\beta V}2\sum_{\sigma\sigma'}f_0^{\sigma\sigma'}\rho_{\sigma}\rho_{\sigma'}
  \nonumber\\~~~~~
  = -\beta V \Big\{\sum_{\sigma}\int_{\bd k}\ln\Big[
  \frac{1+e^{-\beta\tilde{\xi}_{{\bd k}\sigma}}}{1+e^{-\beta\xi_{{\bd k}\sigma}}}\Big]
  + \frac12\sum_{\sigma\sigma'} f_0^{\sigma\sigma'}\rho_{\sigma}\rho_{\sigma'}
  \Big\}\,,\\
  \Gamma^{(0,1)}_{\Lambda_0,\sigma} = -i \left[\rho_{\sigma} - \int_{K}\tilde{G}_{0,\sigma}(K)\right]\,,\\
  \Gamma^{(2,1)}_{\Lambda_0}(K+\bar{K},\sigma;K\sigma;\bar{K}\sigma)=i\,,\\
  \Gamma^{(2,0)}_{\Lambda_0}(K\sigma;K\sigma)=\Sigma^{\Lambda_0}_{\sigma}(K)
  = \sum_{\sigma'}f_0^{\sigma\sigma'}\rho_{\sigma'}\,,\\
  \Gamma^{(0,2)}_{\Lambda_0}(-\bar{K}\sigma,\bar{K}\sigma) = \Pi^{\Lambda_0}_{\sigma,\sigma'}(\bar{K})
  = [f_0^{-1}]^{\sigma\sigma'}\delta_{\bar{K},0}/(\beta V) + L^{(2)}_{\sigma}(-\bar{K},\bar{K})\,,
\end{eqnarray}
and for $m\ge2$,
\begin{eqnarray}
  \Gamma^{(0,m)}(\bar{K}_1\sigma, \dots , \bar{K}_m\sigma)
  = L_{\sigma}^{(m)}(\bar{K}_1,\dots,\bar{K}_m)\,.
\end{eqnarray}
All other one-line irreducible vertices vanish initially. Here, we
have defined the symmetrized fermion loops as
\begin{eqnarray}
  \fl L_{\sigma}^{(n)}(\bar{K}_1,\dots,\bar{K}_n)=
  \frac{i^n}{n}\sum_{P}\int_K \tilde{G}_{0,\sigma}(K)\tilde{G}_{0,\sigma}(K+\bar{K}_{P(1)})
  \nonumber\\
  ~~~~~~~~~~~~~~~~~~~
  \cdot\dots\cdot \tilde{G}_{0,\sigma}(K+\bar{K}_{P(1)}+\ldots+\bar{K}_{P(n-1)})\,.
\end{eqnarray}
The summation is over all permutations $P$ of $\{1,\dots,n\}$.  If we
now demand that $\Gamma_{\Lambda_0}^{(0,1)}=0$, we obtain the relation
\begin{equation}
  \rho_{\sigma} = \int_K \tilde{G}_{0,\sigma}(K) = \int_{\bd k}
  n_F (\xi_{{\bd k}\sigma}+\sum_{\sigma'}f_0^{\sigma\sigma'}\rho_{\sigma'})\,,
  \label{eq:Hartree}
\end{equation}
which is nothing but a Hartree self-consistency condition. Hence, the
initial condition in the interaction-cutoff scheme is simply the
self-consistent Hartree approximation for the density. Note that
Eq.~(\ref{eq:Hartree}) can also have ferromagnetic solutions with
$\rho_{\uparrow}\neq\rho_{\downarrow}$.

By solving the flow equations for the purely bosonic vertices in some
appropriate truncation, our approach then allows for a systematic
inclusion of bosonic fluctuations around the self-consistent Hartree
approximation.  Since the fermionic degrees of freedom are not
completely integrated out in our approach, it is in principle possible
to use this solution for the bosonic vertices in the flow of vertices
with external fermionic legs, e.g. to calculate approximations to the
single-particle Green's function.

\section{Summary and outlook}

We have presented a unified way of treating vacuum expectation values
of bosonic fields in the functional renormalization group.  In a
simple truncation, our flow equations for the Ising model in the
symmetry broken phase reduce to well known results obtained in the
context of the local potential approximation. We apply our formalism
to the coupled Bose-Fermi theory obtained by decoupling the
electron-electron interaction in the zero-sound channel. The vacuum
expectation value of the zero-frequency and zero-momentum mode of the
bosonic field is in this case closely related to the fermionic
density.  The dependence of this vacuum expectation value at the end
of the RG flow on the initial chemical potential allows us to
calculate the compressibility of the system. Using a fermionic cutoff
and a simple Hartree-type truncation of the flow equations, we obtain
the known Fermi-liquid result for the compressibility. In the
interaction-cutoff scheme, a cutoff is introduced solely in the
bosonic propagator and a fermionic determinant has to be evaluated to
determine the initial condition of the flow. This initial condition is
then given by the self-consistent Hartree approximation and the flow
equations can be used to obtain corrections to this mean-field theory.

It is straightforward to generalize the partial bosonization approach
to the case where some other collective bosonic field develops a
finite expectation value. For example, with the help of a
Hubbard-Stratonovich transformation in the particle-particle channel,
we may study fluctuation corrections to the BCS approximation for a
superconductor. Note that within the interaction-cutoff scheme, the
self-consistent BCS approximation should emerge as the initial
condition for the coupled Fermi-Bose theory, in analogy to
Eq.~(\ref{eq:Hartree}).  This should be compared with the purely
fermionic functional RG approach to symmetry breaking developed by
Salmhofer and coauthors \cite{Salmhofer04}, where the BCS
self-consistency condition is only generated in the limit $\Lambda
\rightarrow 0$ by the RG flow starting from an infinitesimal symmetry
breaking perturbation.

The truncation scheme used in Sec.~\ref{sec:Ising} to study symmetry
breaking in the classical Ising model can be applied to any bosonic
sector of quantum systems involving both bosons and fermions.  Such a
procedure, which is equivalent to a truncated local potential
approximation, might be a convenient starting point to study the
feedback between bosonic fluctuations on the fermionic single-particle
excitations.

Finally, it should be mentioned that an alternative functional RG
approach, which also seems to be convenient to study spontaneous
symmetry breaking, has recently been developed by Dupuis
\cite{Dupuis05}.  His approach is based on the functional RG flow of
the generating functional of the two-particle irreducible vertex
functions.  Note that in fermionic language the one-line irreducible
vertex functions introduced in our partial bosonization approach are
only partially two-particle irreducible in the sense that two-particle
irreducible diagrams singled out by the Hubbard-Stratonovich field are
eliminated in favour of a collective bosonic field.

We appreciate useful discussions with N. Hasselmann, L. Bartosch and
S. Ledowski.  This work was partially supported by the DFG via the
Forschergruppe FOR 412.

\appendix

\renewcommand{\theequation}{A.\arabic{equation}}
\renewcommand{\thesubsection}{A.\arabic{subsection}}

\section*{Appendix: Initial condition in the interaction-cutoff scheme}

A more formal route to the initial condition (\ref{eq:Gamma0}) for
$\Gamma$ in the interaction-cutoff scheme uses an additional
generating functional for partially amputated-connected Green's
functions for which external interaction lines are amputated
\cite{Schuetzphd},
\begin{equation}
  \fl e^{\G_{\rm pac}[\bar{\jmath},j,J]}=
  \frac{1}{\Z_0}
  \int D[\bar{\psi},\psi,\varphi] e^{-S_0[\bar{\psi},\psi]-S_0[\varphi]-S_1[\bar{\psi},\psi,\varphi+J]
    +(\bar{\jmath},\psi)+(\bar{\psi},j)}\,.
  \label{eq:Gpacdef}
\end{equation} 
By elementary manipulation of the functional integral, we can show
\begin{equation}
  \fl e^{\G_{\rm pac}[\bar{\jmath},j,J]}=e^{\frac12 \left(\p{J},\hat{F}_0\p{J}\right)}
  \frac1{\Z_{\psi}}\int D[\bar{\psi},\psi] e^{-S_0[\bar{\psi},\psi]-S_1[\bar{\psi},\psi,J]+
    (\bar{\jmath},\psi)+(\bar{\psi},j)}\,.
  \label{eq:Gpacintermediate}
\end{equation}
Here, $[\hat{F}_0]_{\bar{K}\sigma,\bar{K}'\sigma'}=
\delta_{\bar{K}+\bar{K}',0}F_{0,\sigma\sigma'}(\bar{K}')$ is the
matrix of the bare interaction, and $\Z_{\psi}=\int D[\bar{\psi},\psi]
e^{-S_0[\bar{\psi},\psi]}$ is the partition function for
non-interacting particles. For the initial condition of the flow, we
have $f\to0$.  This limit can readily be taken in
Eq.~(\ref{eq:Gpacintermediate}).  The remaining functional integral is
Gaussian and yields
\begin{equation}
  \G_{\rm pac}^{f\to0}[\bar{\jmath},j,J]=
  \Tr \{ \ln [\hat{1}-i\hat{G}_0\hat{J}]\}
  - (\bar{\jmath},[\hat{1}-i\hat{G}_0\hat{J}]^{-1}\hat{G}_0 j)\,,
  \label{eq:Gpacexplicit}
\end{equation}
where we use a matrix $\hat{J}$ of bosonic sources containing the
matrix elements
\begin{equation}
  [\hat{J}]_{K\sigma,K'\sigma'} = \delta_{\sigma\sigma'} J_{K-K',\sigma}\,.
  \label{eq:Jmatrixdef}
\end{equation}
The first term in Eq.~(\ref{eq:Gpacexplicit}) generates the closed
loops of fermion propagators in Fig.~\ref{fig:closedloop} when
expanded in powers of the bosonic sources.  The second term generates
diagrams that contain a continuous fermionic path linking two external
fermionic legs. An arbitrary number of external bosonic legs are then
directly attached to this line. The latter diagrams are not
one-line irreducible and will cancel in the expression for $\Gamma$.
However, before we can perform the Legendre transformation, we first
need a relation between $\G_{\rm pac}$ and the generating functional
$\G_c$ for connected Green's functions. This is achieved by a shift
$\varphi\to\varphi-J$ in the integration variables in
Eq.~(\ref{eq:Gpacdef}) and yields
\begin{equation}
  \G_c[\bar{\jmath},j,J]=S_0[\tilde{J}] + \G_{\rm pac}[\bar{\jmath},j,\tilde{J}]\,,
  \label{eq:GcGpacrelation}
\end{equation}
where we have defined $\tilde{J}=\hat{F}_0 J$. The classical field
$\varphi$ is then given by
\begin{equation}
  \varphi = \pp{\G_c}{J} = \tilde{J} + \hat{F}_0\pp{\G_{\rm pac}}{\tilde{J}}[\bar{\jmath},j,\tilde{J}]
  \stackrel{f\to0}{=} \tilde{J}\,.
  \label{eq:varphifto0}
\end{equation}
In the limit of a vanishing interaction the classical field $\varphi$
thus becomes identical to the source field $\tilde{J}$. Since
$\G_{\mathrm{pac}}^{f\to0}$ is quadratic in the fermionic sources, the
inversion necessary to obtain the sources $\bar{\jmath}$ and $j$ as a
function of the classical fields involves just a matrix inversion.
The remaining Legendre transformation can then be explicitly performed
and we obtain Eq.~(\ref{eq:Gamma0}).

\end{document}